\documentclass[
  reprint,
  superscriptaddress,
  amsmath,
  amssymb,
  aps,
  floatfix,
]{revtex4-2}

\usepackage{graphicx}
\graphicspath{images/}
\usepackage{hyperref}
\usepackage{float}

\newcommand{\proton}{p}
\newcommand{\pion}{\pi^{-}}
\newcommand{\sNN}{\sqrt{s_\mathrm{NN}}}
\newcommand{\Jsys}{\vec{J}}
\newcommand{\JsysHat}{\hat{J}}
\newcommand{\REP}{R_\mathrm{EP}^{(1)}}
\newcommand{\pT}{p_\mathrm{T}}
\newcommand{\InvMass}{m_\mathrm{inv}}
\newcommand{\pstar}{_\proton^{*}}
\newcommand{\PpStar}{\vec{p}_\proton^{\,*}} 
\newcommand{\PsiEP}{\Psi_{1}}
\newcommand{\PsiEpd}{\Psi_{1, \mathrm{EPD}}}
\newcommand{\PLambda}{\overline{P}_{\Lambda}}
\newcommand{\PLamBar}{\overline{P}_{\bar{\Lambda}}}
\newcommand{\PHyper}{\overline{P}_\mathrm{H}}
\newcommand{\sig}{^\mathrm{sig}}
\newcommand{\bg}{^\mathrm{bg}}
\newcommand{\obs}{^\mathrm{obs}}
\newcommand{\BSTAR}{\vec{B}_\mathrm{STAR}}
\newcommand{\AzimuthalEmissionAngle}{\phi_\Lambda-\phi\pstar}
\newcommand{\PolarizationCorrelationTerm}{\left<\sin(\PsiEP-\phi\pstar)\right>}
\newcommand{\LambdaPdgMass}{m_{\Lambda, \mathrm{PDG}}}
\newcommand{\Efficiency}{\varepsilon\left(y,p_\mathrm{T}\right)}
\newcommand{\AZero}{\tfrac{4}{\pi}\overline{\sin\theta\pstar}}
\newcommand{\RightLeftTripleProduct}{\left(\vec{p}_\Lambda\times\PpStar\right)\cdot\BSTAR}

\begin{document}

\preprint{APS/123-QED}

\title{Global \texorpdfstring{$\Lambda$}{Λ}-hyperon polarization in Au+Au collisions at \texorpdfstring{$\sNN=3$}{√sNN=3} GeV}

\affiliation{Abilene Christian University, Abilene, Texas   79699}
\affiliation{AGH University of Science and Technology, FPACS, Cracow 30-059, Poland}
\affiliation{Alikhanov Institute for Theoretical and Experimental Physics NRC "Kurchatov Institute", Moscow 117218, Russia}
\affiliation{Argonne National Laboratory, Argonne, Illinois 60439}
\affiliation{American University of Cairo, New Cairo 11835, New Cairo, Egypt}
\affiliation{Brookhaven National Laboratory, Upton, New York 11973}
\affiliation{University of California, Berkeley, California 94720}
\affiliation{University of California, Davis, California 95616}
\affiliation{University of California, Los Angeles, California 90095}
\affiliation{University of California, Riverside, California 92521}
\affiliation{Central China Normal University, Wuhan, Hubei 430079 }
\affiliation{University of Illinois at Chicago, Chicago, Illinois 60607}
\affiliation{Creighton University, Omaha, Nebraska 68178}
\affiliation{Czech Technical University in Prague, FNSPE, Prague 115 19, Czech Republic}
\affiliation{Technische Universit\"at Darmstadt, Darmstadt 64289, Germany}
\affiliation{ELTE E\"otv\"os Lor\'and University, Budapest, Hungary H-1117}
\affiliation{Frankfurt Institute for Advanced Studies FIAS, Frankfurt 60438, Germany}
\affiliation{Fudan University, Shanghai, 200433 }
\affiliation{University of Heidelberg, Heidelberg 69120, Germany }
\affiliation{University of Houston, Houston, Texas 77204}
\affiliation{Huzhou University, Huzhou, Zhejiang  313000}
\affiliation{Indian Institute of Science Education and Research (IISER), Berhampur 760010, India}
\affiliation{Indian Institute of Science Education and Research (IISER) Tirupati, Tirupati 517507, India}
\affiliation{Indian Institute Technology, Patna, Bihar 801106, India}
\affiliation{Indiana University, Bloomington, Indiana 47408}
\affiliation{Institute of Modern Physics, Chinese Academy of Sciences, Lanzhou, Gansu 730000 }
\affiliation{University of Jammu, Jammu 180001, India}
\affiliation{Joint Institute for Nuclear Research, Dubna 141 980, Russia}
\affiliation{Kent State University, Kent, Ohio 44242}
\affiliation{University of Kentucky, Lexington, Kentucky 40506-0055}
\affiliation{Lawrence Berkeley National Laboratory, Berkeley, California 94720}
\affiliation{Lehigh University, Bethlehem, Pennsylvania 18015}
\affiliation{Max-Planck-Institut f\"ur Physik, Munich 80805, Germany}
\affiliation{Michigan State University, East Lansing, Michigan 48824}
\affiliation{National Research Nuclear University MEPhI, Moscow 115409, Russia}
\affiliation{National Institute of Science Education and Research, HBNI, Jatni 752050, India}
\affiliation{National Cheng Kung University, Tainan 70101 }
\affiliation{Nuclear Physics Institute of the CAS, Rez 250 68, Czech Republic}
\affiliation{Ohio State University, Columbus, Ohio 43210}
\affiliation{Institute of Nuclear Physics PAN, Cracow 31-342, Poland}
\affiliation{Panjab University, Chandigarh 160014, India}
\affiliation{Pennsylvania State University, University Park, Pennsylvania 16802}
\affiliation{NRC "Kurchatov Institute", Institute of High Energy Physics, Protvino 142281, Russia}
\affiliation{Purdue University, West Lafayette, Indiana 47907}
\affiliation{Rice University, Houston, Texas 77251}
\affiliation{Rutgers University, Piscataway, New Jersey 08854}
\affiliation{Universidade de S\~ao Paulo, S\~ao Paulo, Brazil 05314-970}
\affiliation{University of Science and Technology of China, Hefei, Anhui 230026}
\affiliation{Shandong University, Qingdao, Shandong 266237}
\affiliation{Shanghai Institute of Applied Physics, Chinese Academy of Sciences, Shanghai 201800}
\affiliation{Southern Connecticut State University, New Haven, Connecticut 06515}
\affiliation{State University of New York, Stony Brook, New York 11794}
\affiliation{Instituto de Alta Investigaci\'on, Universidad de Tarapac\'a, Arica 1000000, Chile}
\affiliation{Temple University, Philadelphia, Pennsylvania 19122}
\affiliation{Texas A\&M University, College Station, Texas 77843}
\affiliation{University of Texas, Austin, Texas 78712}
\affiliation{Tsinghua University, Beijing 100084}
\affiliation{University of Tsukuba, Tsukuba, Ibaraki 305-8571, Japan}
\affiliation{Valparaiso University, Valparaiso, Indiana 46383}
\affiliation{Variable Energy Cyclotron Centre, Kolkata 700064, India}
\affiliation{Warsaw University of Technology, Warsaw 00-661, Poland}
\affiliation{Wayne State University, Detroit, Michigan 48201}
\affiliation{Yale University, New Haven, Connecticut 06520}

\author{M.~S.~Abdallah}\affiliation{American University of Cairo, New Cairo 11835, New Cairo, Egypt}
\author{B.~E.~Aboona}\affiliation{Texas A\&M University, College Station, Texas 77843}
\author{J.~Adam}\affiliation{Brookhaven National Laboratory, Upton, New York 11973}
\author{L.~Adamczyk}\affiliation{AGH University of Science and Technology, FPACS, Cracow 30-059, Poland}
\author{J.~R.~Adams}\affiliation{Ohio State University, Columbus, Ohio 43210}
\author{J.~K.~Adkins}\affiliation{University of Kentucky, Lexington, Kentucky 40506-0055}
\author{G.~Agakishiev}\affiliation{Joint Institute for Nuclear Research, Dubna 141 980, Russia}
\author{I.~Aggarwal}\affiliation{Panjab University, Chandigarh 160014, India}
\author{M.~M.~Aggarwal}\affiliation{Panjab University, Chandigarh 160014, India}
\author{Z.~Ahammed}\affiliation{Variable Energy Cyclotron Centre, Kolkata 700064, India}
% ...

\collaboration{STAR Collaboration}\noaffiliation

\date{\today}

\begin{abstract}
Global hyperon polarization, $\PHyper$, in Au+Au
 collisions over a large range of
 collision energy, $\sNN$, was recently
 measured and successfully reproduced by hydrodynamic
 and transport models with intense fluid
 vorticity of the quark-gluon plasma.
  While na\"{i}ve extrapolation of data
 trends suggests a large $\PHyper$ as
 the collision energy is reduced, the
 behavior of $\PHyper$ at small $\sNN<7.7$~GeV
 is unknown. Operating the STAR experiment
 in fixed-target mode, we measured
 the polarization of $\Lambda$ hyperons along
 the direction of global angular momentum
 in Au+Au collisions at $\sNN=3$~GeV.
  The observation of substantial polarization
 of $4.91\pm0.81(\rm stat.)\pm0.15(\rm syst.)$\% in these
 collisions may require a reexamination of
 the viscosity of any fluid created
 in the collision, of the thermalization timescale
 of rotational modes, and of hadronic
 mechanisms to produce global polarization.
\end{abstract}

\maketitle

Collisions between heavy nuclei at the
 highest energies at the Relativistic Heavy
 Ion Collider (RHIC) and the Large
 Hadron Collider (LHC) produce the quark-gluon
 plasma (QGP), a strongly interacting system
 characterized by colored degrees of freedom\cite{Akiba:2015jwa}.
 Viscous relativistic hydrodynamics is one of
 the most powerful tools to understand
 this system theoretically\cite{Teaney:2009qa}; it is the
 dynamical heart of the ``standard model
 of the Little Bang"\cite{Heinz:2013wva}. Systematic comparisons
 of data and the hydrodynamic response
 to anisotropies in the initial state
 have yielded considerable insight on transport
 coefficients and the equation of state
 of the QGP\cite{Bernhard:2016tnd}. Recently, considerable experimental
 and theoretical efforts have focused on
 the polarization of particles emitted from
 the fluid\cite{Becattini:2020ngo}---mostly $\Lambda$ hyperons\cite{Liang:2004ph,Becattini:2007sr,Betz:2007kg,Abelev:2007zk,STAR:2017ckg,Acharya:2019ryw} and,
 very recently, multi-strange hyperons\cite{Adam:2020pti}---which probe
 the local vorticity of the fluid.

Hydrodynamic\cite{Becattini:2020ngo} and transport\cite{Vitiuk:2019rfv,Li:2017slc} simulations each reproduce
 rather well the measured ``global'' polarization,
 the component directed along the total
 angular momentum of the collision, $\Jsys$.
 In most hydrodynamic calculations, particle properties
 (e.g., momentum and flavor) are derived
 from fluid properties (e.g., stress-energy tensor
 and chemical potentials) through the Cooper-Frye
 ansatz\cite{Becattini:2020ngo}, which assumes equilibrium at the
 point of hadronization. This formalism has
 been generalized\cite{Becattini:2013fla} to calculate particle polarization
 directly from the thermal vorticity\cite{Becattini:2020ngo} of
 the fluid. Equilibration of orbital angular
 momentum and spin degrees of freedom
 is assumed, though spin relaxation times
 are not fully understood\cite{Bhadury:2020puc}. In transport
 simulations, the vorticity is calculated from
 the particles in small cells, and
 polarization is extracted through the generalized
 Cooper-Frye formalism discussed above.

The first measurement of $\PHyper$ by the STAR
 Collaboration at $\sNN=62.4$ and $200$~GeV
 was consistent with zero\cite{Abelev:2007zk}; however, subsequent
 measurements across a range of lower
 collision energies $7.7\leq\sNN\leq39$~GeV and with higher
 statistics at $\sNN=200$~GeV by the STAR
 Collaboration showed statistically significant $\PHyper>0$\cite{STAR:2017ckg,Adam:2018ivw}. Together
 with high-statistics measurements at $\sNN=2.76$ and
 $5.02$~TeV by the ALICE Collaboration showing
 $\PHyper$ consistent with zero, these measurements
 demonstrated a rising $\PHyper$ with decreasing
 $\sNN$\cite{STAR:2017ckg,Adam:2018ivw,Acharya:2019ryw}. 

While a simple extrapolation of this
 trend would suggest that $\PHyper$ continues
 to rise as $\sNN$ decreases, we
 expect vanishing $\PHyper$ at $\sNN=2m_\mathrm{N}$ due
 to the lack of system angular
 momentum\cite{Deng:2020ygd}. A peak $\PHyper$ therefore likely
 exists in the region $2m_{\rm N}\approx1.9<\sNN<7.7$~GeV;
 recent model calculations predict this peak
 in the vicinity of $\sNN\approx3$~GeV\cite{Deng:2020ygd,Ivanov:2019ern,guo2021locating}. Furthermore,
 these calculations, which at $\sNN\gtrsim7.7$~GeV agree
 fairly well with each other and
 with other higher-$\sNN$ calculations\cite{Li:2017slc,Vitiuk:2019rfv,Sun:2017xhx,Ivanov:2019ern,Karpenko:2016jyx}, diverge for
 $\sNN\lesssim7.7$~GeV. Measurements of $\PHyper$ at $\sNN<7.7$~GeV
 will provide constraints on which sets
 of assumptions are valid at such
 small $\sNN$.

As in previous studies, $\PHyper$ represents
 the spin polarizations of $\Lambda$
 and $\bar{\Lambda}$ hyperons, $\PLambda$ and $\PLamBar$;
 however, $\bar{\Lambda}$-hyperon yields at $\sNN=3$~GeV are
 insufficient for a meaningful study
 of $\PLamBar$ and we therefore refer directly to $\PLambda$.
 We report in this work our
 observation of nonzero $\PLambda$, with a
 statistical significance of nearly $6\sigma$. This
 observation raises important questions:
 What is the spin equilibration timescale,
 and how does it compare to
 the thermal equilibration timescale? How viscous
 is the region of nuclei overlap?
 Our observation of significant, nonzero global
 $\PLambda$ at $\sNN=3$~GeV is the
 largest $\PLambda$ yet observed and the
 lowest energy at which $\PLambda$ has
 been measured.

The dataset discussed in this work
 was collected in 2018 by the
 STAR experiment\cite{Ackermann:2002ad}. The STAR detector configuration
 features the cylindrical geometry characteristic of
 collider experiments. In order to explore
 various regions of the QCD phase
 diagram, RHIC has undertaken a multiyear
 Beam Energy Scan\cite{Aggarwal:2010cw} program, extending observations
 to lower energies. While the maximum
 energy of a gold beam in
 the RHIC ring is 100~GeV per
 nucleon, the facility is remarkably flexible
 and beams with energy as low
 as 3.85~GeV per nucleon can be
 maintained for reasonable times; thus, the
 lowest energy measured in beam-on-beam collisions
 is $\sNN=7.7$~GeV. However, operating the facility
 and experiment in fixed-target mode, in
 which the beam collides with a
 foil target inside the beam pipe
 positioned 200 cm away from the center
 of the Time Projection Chamber (TPC), produces collisions at
 energies as low as $\sNN=3$~GeV. See Ref. \cite{STAR:2020dav}
 for details of the STAR fixed-target configuration.

Charged-particle tracks in the pseudorapidity range $-2\lesssim\eta\lesssim0$
 are measured in the TPC\cite{Anderson:2003ur}. For
 $-1.5\lesssim\eta\lesssim0$, additional identification is performed by
 time-of-flight measurements in the Barrel Time-of-Flight
 (BTOF) detector\cite{Llope:2012zz,Shao:2005iu}. At
 $\eta<-2.55$, charged particles are registered
 in the Event Plane Detector (EPD)\cite{Adams:2019fpo}.
 Pseudorapidity is reported 
 in the laboratory frame while rapidity, $y$, is reported in the
 collision center-of-momentum frame, boosted by the
 beam rapidity, $y_{\rm beam}=1.045$.
 After basic offline selections
 to ensure that the reconstructed collision
 occurred in the target foil, $253\times10^6$
 events were available for this analysis.

The centrality of an event, which
 describes the degree to which the
 colliding nuclei overlap, was estimated based
 on the number of ``primary'' tracks,
 which are mainly determined by checking
 if a track's helical path comes
 within 3~cm of the primary vertex.
 Fitting this multiplicity distribution to a
 Monte Carlo Glauber model\cite{Ray:2007av} calculation provided a
 measure of the centrality and an
 estimate of the trigger efficiency. Details
 of the Glauber calculation for STAR
 fixed-target measurements are given in Ref. \cite{STAR:2020dav}.

Protons and pions measured in the
 TPC were used to reconstruct $\Lambda$
 hyperons, which decay via $\Lambda\rightarrow \proton+\pion$
 63.9\% of the time\cite{Zyla:2020zbs}.  Hyperon
 candidates were constructed through decay topology
 with quality assurance selections, such as
 the distance of closest approach (DCA)
 between the reconstructed $\Lambda$-hyperon trajectory and
 the primary vertex, and the DCA
 between the two daughters.  Details
 on the $\Lambda$-hyperon reconstruction may be
 found in Refs. \cite{Adam:2018ivw,GorbunovThesis,MaksymThesis}.  Additional important kinematic
 selection criteria, or ``cuts", include $\pT>0.4~{\rm GeV}/c$ on daughter protons, to avoid
 contamination with spallation particles, and $\pT>0.15~{\rm GeV}/c$ on daughter pions, as tracking
 efficiency in STAR drops quickly for
 lower values of transverse momentum. 
 Finally, we select $\Lambda$ hyperons with
 $\pT>0.7~{\rm GeV}/c$ as the checks for
 unknown systematic effects fail below $\pT=0.7~{\rm GeV}/c$,
 and reconstruction efficiency on $\Lambda$ hyperons
 also becomes very small below $\pT=0.7~{\rm GeV}/c$. 
 Our coverage in rapidity
 was in the range $-0.2<y<1.0$.
 The typical ratio
 of true $\Lambda$-hyperon yield to combinatoric
 background near $\InvMass=\LambdaPdgMass$ is 10:1.

The parity-violating nature of hyperon decay
 reveals the spin polarization, $\vec{P}_{\Lambda}$. The
 global polarization is the projection of
 $\vec{P}_{\Lambda}$ along the direction of the
 angular momentum of the collision, $\JsysHat$\cite{Abelev:2007zk,Becattini:2020ngo};
 in the case of a symmetric
 collision system and detector setup we
 may write this as\cite{STAR:2017ckg}
\begin{equation}
\label{eq:GlobalPol}
  \PLambda \equiv \left\langle\vec{P}_{\Lambda}\cdot\JsysHat\right\rangle 
  = \frac{8}{\pi\alpha_{\Lambda}}\frac{1}{\REP}\left\langle\sin\left(\PsiEP-\phi\pstar\right)\right\rangle\sig.
\end{equation}
Here, $\alpha_\Lambda=0.732\pm0.014$ is the $\Lambda$-hyperon decay
 constant\cite{Zyla:2020zbs}, $\phi\pstar$ is the azimuthal angle
 of the daughter proton momentum in
 the $\Lambda$-hyperon rest frame, and the
 average is over all hyperons in
 the momentum range selected. $\PsiEP$ is
 the first-order event plane of the
 collision, and $\REP$ is the resolution
 with which the event plane estimates
 the reaction plane of the collision,
 which is normal to $\JsysHat$. The
 ``sig'' label indicates that the average
 excludes combinatoric-background contributions and acceptance effects;
 we discuss these below. As we
 will see later, the symmetries required
 to achieve the form of Eq.
 (\ref{eq:GlobalPol}) are broken when operating RHIC/STAR
 in fixed-target mode.

Charged particles with $-2.84<\eta<-2.55$, measured in
 the outer four rings of the
 EPD, are used to determine $\PsiEP$\cite{Adams:2019fpo},
 and the three-subevent method\cite{Poskanzer:1998yz} is used
 to measure $\REP$. The two reference
 subevents used in this method use
 particles measured at $-0.5<\eta<-0.4$ and $-0.2<\eta<-0.1$
 in the TPC. In this analysis,
 the event-plane resolution $\REP\approx40\%$ for 20--50\%
 central collisions. Because the STAR magnetic
 field along $\hat{z}$ causes charged particles
 to curve and also because produced
 particles are disproportionately positive as $\sNN$
 becomes smaller, $\PsiEP$ as measured by
 the EPD is twisted by an
 angle $\Delta\PsiEpd$. $\Delta\PsiEpd$ can be calculated
 by correlating $\PsiEpd$ with $\Psi_{1, \rm{TPC}}$,
 as the TPC is able to
 trace tracks to the collision point
 and therefore does not suffer this
 rotation effect; $\Delta\PsiEpd=0.063\pm0.011$ by which we correct $\PsiEpd$.

A fraction of $[\proton,\pion]$ pairs that
 enter our analysis will not arise
 from true hyperons, but will instead
 originate from combinatorial background. To statistically
 extract the true polarization signal from
 the false background signal, we used
 the invariant-mass method\cite{Adam:2018ivw,Adamczyk:2013gw,Borghini:2004ra}, in which the
 observed $\left\langle\sin\left(\PsiEP-\phi\pstar\right)\right\rangle$ is measured as a
 function of invariant mass and written
 as a sum of signal and
 background contributions:
\begin{align}
\label{eq:InvMassMethod}
  \langle\sin(\PsiEP&-\phi\pstar)\rangle\obs\left(\InvMass\right)\\
  =&f\bg\left(\InvMass\right)\PolarizationCorrelationTerm\bg \nonumber \\
  +&\left(1-f\bg\left(\InvMass\right)\right)\PolarizationCorrelationTerm\sig. \nonumber
\end{align}
Here, $\PolarizationCorrelationTerm\sig$ is the average $\Lambda$-hyperon
 polarization, while the term $\PolarizationCorrelationTerm\bg$ is
 the false polarization of the combinatoric
 background. The combinatoric fraction $f\bg(\InvMass)$ is
 extracted through fits to the $\InvMass$
 distribution.

\begin{figure}[!t]
  \centering
  \includegraphics[width=\linewidth]{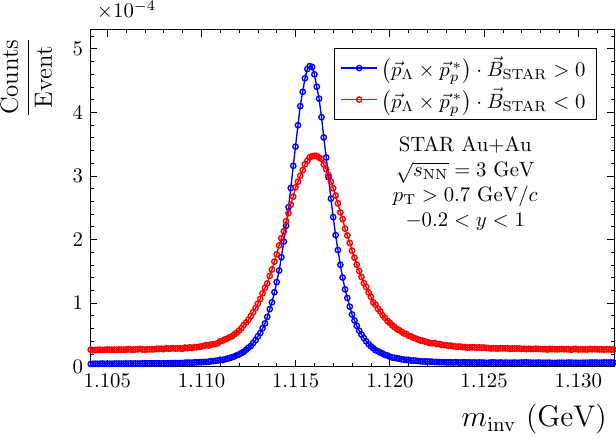}
  \caption{The measured $\InvMass$ distributions of
 two classes of $\Lambda$-hyperon decays: ``right"
 decays in blue, with $\RightLeftTripleProduct>0$, and
 ``left" decays in red, with $\RightLeftTripleProduct<0$.
 The ``right" decay class has a
 notably sharper $\InvMass$ distribution than the
 ``left" decay class, and this is
 due to the effects of daughter
 tracks crossing in the STAR TPC
 with $\BSTAR||-\hat{z}$. The opposite pattern is
 obtained by flipping the sign of
 $\BSTAR$ or by reconstructing $\bar{\Lambda}$ hyperons.}
  \label{fig:InvMassDistributionWidths}
\end{figure}

The direction of the STAR magnetic
 field, $\BSTAR$, which is aligned with
 the direction of the beam momentum
 ($-\hat{z}$) in the laboratory frame, drives
 charged particles to follow helical paths
 and breaks a right-left symmetry in
 the $\Lambda$-hyperon decay. Consider a ``right"
 and a ``left" class of decays.
 A ``right" decay is one in
 which the proton decays to the
 right side of the $\Lambda$ hyperon
 as viewed along $-\hat{z}$, or equivalently
 when $\RightLeftTripleProduct>0$. A ``left" decay simply
 flips the sign of $\PpStar$. Due
 to their helical paths, the tracks
 of daughters from ``left" decays diverge
 while those from ``right" decays cross
 paths in the transverse plane. STAR's
 $\Lambda$-hyperon reconstruction efficiency, resolution, and purity
 therefore depend on $\RightLeftTripleProduct$, leading to
 the differences in the invariant-mass spectra
 shown in Fig.~\ref{fig:InvMassDistributionWidths}. Directed flow, $v_1$,
 modulates the yield of $\Lambda$
 hyperons as $\sim\left(1+v_1\cos\left(\phi_\Lambda-\PsiEP\right)\right)$~\cite{Voloshin:2008dg} and in fixed-target
 mode our acceptance is greater for
 $y>0$ than for $y<0$; there is
 therefore a net directed flow 
 when integrating over all $\Lambda$ hyperons
 such that $\phi_\Lambda$ is positively correlated
 with $\PsiEP$. 
  
\begin{figure}
  \centering
  \includegraphics[width=\linewidth]{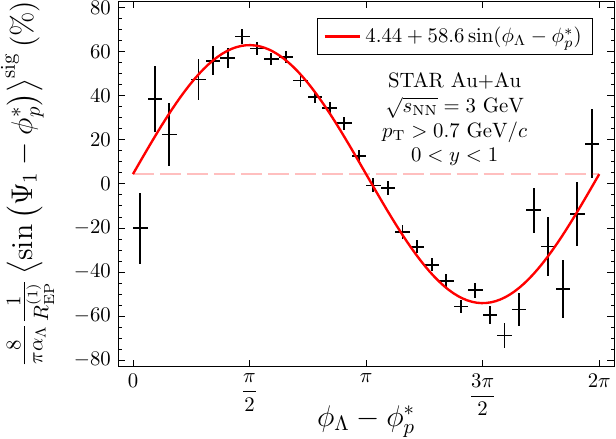}
  \caption{The signal polarizations extracted according
 to Eq. (\ref{eq:InvMassMethod}) as a function of
 $\AzimuthalEmissionAngle$, for positive-rapidity $\Lambda$ hyperons. The
 sinusoidal behavior is driven by nonzero
 net $v_1$. The vertical shift corresponds
 to the vorticity-driven polarization according to
 Eq. (\ref{eq:SineDependence}); in collider mode, where the
 net $v_1$ is zero, this dependence
 on $\AzimuthalEmissionAngle$ does not exist.
 The functional form of Eq. (\ref{eq:SineDependence}) fits the data well, 
 with $\chi^2/\mathrm{ndf}=1.49$.}
  \label{fig:SignalPolarizationVsEmissionAngle}
\end{figure}

Recall the polarization correlator, $\PolarizationCorrelationTerm$, from
 Eq. (\ref{eq:InvMassMethod}); because of the net, flow-driven
 correlation between $\phi_\Lambda$ and $\PsiEP$, the
 aforementioned ``right" (``left") decay will correspond
 to $\PolarizationCorrelationTerm>0\:(<0)$. Since ``left" decays also
 have a wider $\InvMass$ distribution, they
 dominate the sides of the net
 $\InvMass$ distribution while ``right" decays dominate
 the center. The observed net polarization
 correlation term from Eq. (\ref{eq:InvMassMethod}) is therefore
 sharply peaked and positive for $\InvMass\approx\LambdaPdgMass$,
 and becomes negative as $|\InvMass-\LambdaPdgMass|$ becomes
 larger, and therefore does not follow
 the form of the observed net
 $f\sig\left(\InvMass\right)$. For this reason, we generalize
 the invariant-mass method by performing the
 method separately for narrow bins in
 $\AzimuthalEmissionAngle$.
 By expanding the correlator $\PolarizationCorrelationTerm$ 
 and taking advantage of the fact that $\left<\sin(\PsiEP-\phi_\Lambda)\right>=0$
 through symmetry, each bin in $\AzimuthalEmissionAngle$ has a contribution to $\PolarizationCorrelationTerm$
 proportional to $\left<\sin(\phi_\Lambda-\phi\pstar)\right>$.
 Across all bins in $\AzimuthalEmissionAngle$
 the net, flow-driven correlation between
 $\phi_\Lambda$ and $\PsiEP$, present in our
 data, therefore generates a sinusoidal component in
 Eq. (\ref{eq:GlobalPol}) unrelated to global polarization, so that
\begin{equation}
\label{eq:SineDependence}
  \frac{8}{\pi\alpha_\Lambda}\frac{1}{\REP}\left\langle\sin\left(\PsiEP-\phi\pstar\right)\right\rangle\sig = \PLambda  +c\sin(\AzimuthalEmissionAngle),
\end{equation}
where the coefficient $c$ depends on
 $v_1$. Figure~\ref{fig:SignalPolarizationVsEmissionAngle} shows the signal polarizations
 extracted using Eq. (\ref{eq:InvMassMethod}) across small bins
 in $\AzimuthalEmissionAngle$ and fitted according to
 Eq. (\ref{eq:SineDependence}). The vertical shift corresponds to
 $\PLambda$; this procedure removes any contributions
 from potentially nonzero polarization in the
 production plane, spanned by $\vec{p}_\Lambda\times\vec{p}_\mathrm{beam}$, as
 seen in Refs. \cite{Lesnik:1975my,Bunce:1976yb}. This procedure is performed
 separately for $y_\Lambda>0$ and $y_\Lambda<0$, and
 the weighted average is extracted. We
 performed detailed simulations of the
 STAR acceptance and tracking reconstruction to
 verify the above procedure to extract
 $\PLambda$. Previous analyses\cite{Abelev:2007zk,STAR:2017ckg,Adam:2018ivw,Acharya:2019ryw} have focused on
 particles measured near mid-rapidity ($|y|<1$) and
 at higher collision energies, where directed
 flow\cite{Poskanzer:1998yz} is small; as well, the
 $\Lambda$-hyperon acceptance is symmetric in $y$
 in collider mode. The azimuthal dependencies
 discussed above were therefore not an issue.

Finite detector acceptance and efficiency necessitate
 two additional corrections on the measured
 polarization. Equation (\ref{eq:GlobalPol}) assumes that the efficiency
 to measure daughter protons is independent
 of $\PpStar$, the daughter momentum direction
 in the hyperon rest frame. However,
 rapidity cuts and inefficiencies introduce a
 weak dependence on $\PpStar$, leading to
 a correction factor $\AZero$\cite{Abelev:2007zk} which depends
 on $\pT, y$, and centrality and
 is $\mathcal{O}(1\%)$. Similarly, the $\Lambda$-hyperon detection
 efficiency, $\Efficiency$, depends on $\vec{p}_\Lambda$.

\begin{figure}[t]
  \centering
  \includegraphics[width=\linewidth]{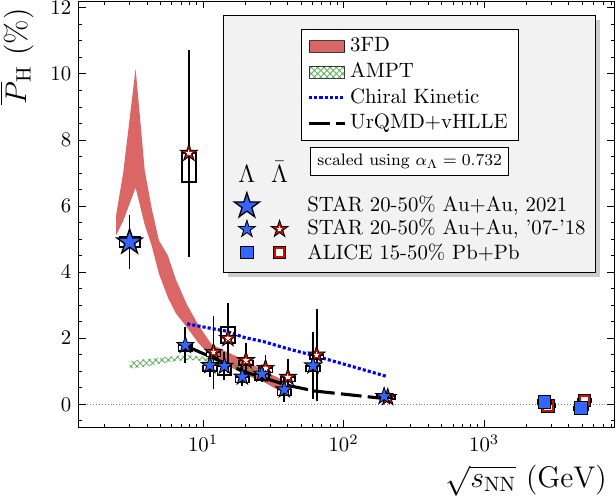}
  \caption{Global hyperon polarization as a
 function of $\sNN$ in mid-central heavy-ion
 collisions. The trend of increasing $\PHyper$
 with decreasing $\sNN$ is maintained at
 the low energy of $\sNN=3$~GeV.
 Statistical uncertainties are represented with lines
 while systematic uncertainties are represented with
 boxes. Previous experimental results\cite{Abelev:2007zk,STAR:2017ckg,Adam:2018ivw,Acharya:2019ryw} are scaled\cite{Becattini:2020ngo}
 using the currently accepted\cite{Zyla:2020zbs} decay parameter $\alpha_\Lambda=0.732$.
 Calculations with a hybrid model (UrQMD+vHLLE)\cite{Karpenko:2016jyx}
 and chiral-kinetic transport\cite{Sun:2017xhx} are compared to
 the higher-$\sNN$ data only, while others
 have been extended to lower energy.
 The AMPT model\cite{guo2021locating} matches higher-energy data
 well while dramatically underestimating $\PLambda$ at
 $\sNN=3$~GeV. The hydrodynamic 3FD model\cite{Ivanov:2020udj} with
 two separate equations of state (crossover
 and first-order phase transition) predicts a
 sharply rising $\PLambda$ below $\sNN=7.7$~GeV.
 The models shown use an impact parameter of 8~fm. }
  \label{fig:PolarVsEnergy}
\end{figure}

A suite of tests was performed
 to search for unexpected systematic effects~\cite{Barlow:2002yb}.
 This included analyzing collisions measured at
 different times during the experiment, checking
 both time of day and day
 of the week; restricting the analysis
 to various regions of $\phi_\Lambda$ in the laboratory system;
  separately analyzing collisions recorded when the collision rate was high or low,
  or with high or low
 experimental background rates; changing the $\Lambda$-finding
 algorithm; changing the numerous fit parameters
 in the invariant-mass method; changing the
 width in $\eta$ of the subevent
 used for $\PsiEP$ calculation; and changing
 the set of topological cuts used to identify $\Lambda$ hyperons. 
Contributions to systematic uncertainty originate in
 the uncertainties on our measurements of
 the corrections.
 These contributions include a 2\% systematic uncertainty
  is associated with the uncertainty~\cite{Zyla:2020zbs} on
  $\alpha_\Lambda$; a $<1$\% statistical uncertainty on
  $\Efficiency$ corresponding to the statistical precision
  of the Monte Carlo simulations; a
  $<1$\% statistical uncertainty on $\AZero$;
  a $<1$\% uncertainty on $\Delta\PsiEpd$; a
  $<1$\% statistical uncertainty on $\REP$; and
  a $<1$\% uncertainty arising from the
  assumptions made about the background polarization's
  dependence on $\InvMass$ when applying Eq. (\ref{eq:InvMassMethod}).
 These systematic
 uncertainties are added in quadrature to
 get the full systematic uncertainty.

\begin{figure}[t]
  \centering
  \includegraphics[width=\linewidth]{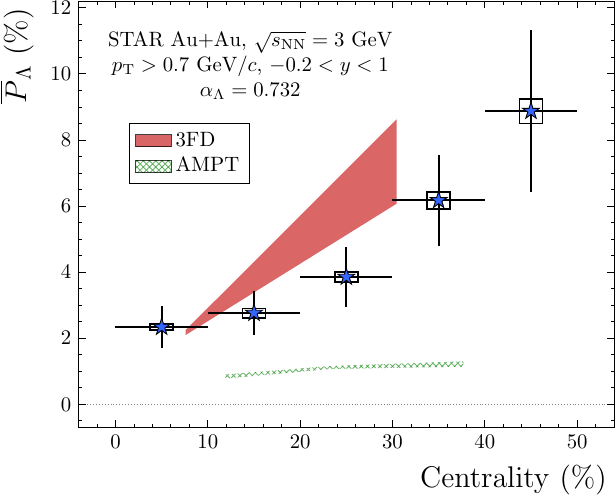}
  \caption{The centrality dependence of $\PLambda$ 
           is compared with partonic-transport\cite{guo2021locating} 
           and three-fluid hydrodynamic\cite{Ivanov:2020udj} calculations at
           impact parameters of 8~fm. Statistical
           uncertainties are represented with lines while
           systematic uncertainties are represented with boxes.
            \label{fig:Centrality}}
\end{figure}

Figure~\ref{fig:PolarVsEnergy} shows the global polarization at
 mid-rapidity, alongside previous measurements whose data
 points have been scaled according to
 the updated decay parameter $\alpha_\Lambda=0.732$\cite{Zyla:2020zbs}. The
 polarization of $\PLambda=4.91\pm0.81(\rm stat.)\pm0.15(\rm syst.)\%$, reported
 in this paper, is the largest
 global $\Lambda$-hyperon polarization yet observed. We
 find that the steady increase of
 $\PLambda$ with decreasing $\sNN$ continues almost
 to the $\Lambda$-hyperon production threshold.

Nevertheless, this trend has been reproduced
 by hydrodynamic and transport calculations\cite{Becattini:2020ngo,Vitiuk:2019rfv,Li:2017slc,Sun:2017xhx} above
 $\sNN=7.7$~GeV. Vorticity from the three-fluid hydrodynamics
 (3FD)\cite{Ivanov:2020udj} as well as
 partonic-transport (AMPT)\cite{guo2021locating} calculations
 have been extended to the lowest
 energies, as shown in Fig.~\ref{fig:PolarVsEnergy}.
 For the hydrodynamic 3FD calculation,
 $\vec{\omega}_{\rm th}$ is calculated directly from
 the local flow and temperature distributions.
 In the AMPT calculations, the thermal
 vorticity is calculated in coarse-grained ``cells''
 from particle ensembles\cite{Li:2017slc}.

Polarizations predicted by 3FD calculations depend
 on the range of hydrodynamic rapidity
 $y_\mathrm{h}\equiv\ln\left[\left(u_0+u_z\right)/\left(u_0-u_z\right)\right]$ of the fluid contributing to
 the $\Lambda$ hyperons\cite{Ivanov:2020wak}. The shaded band
 representing the 3FD model in Fig.~\ref{fig:PolarVsEnergy}
 corresponds to varying the selection between
 $|y_\mathrm{h}|<0.35$ and $|y_\mathrm{h}|<0.6$. Calculations were performed
 using one equation of state in
 which the deconfinement transition is characterized
 as first order and using another
 assuming a crossover transition; the resulting
 difference in polarization between these two
 methods is much smaller than the
 width of the band.

We find that, while the central value of the 
 3FD calculation\cite{Ivanov:2020udj}
 overshoots the measurement
 at $\sNN=3$~GeV by $\sim30\%$, 
 the prediction and our measurement 
 roughly agree within uncertainties.
 The partonic-transport
 calculation\cite{guo2021locating}, which reproduces the measurements quite
 well at $\sNN\geq7.7$~GeV, dramatically underestimates $\PLambda$
 at $\sNN=3$~GeV; the model was tuned
 for very low collision energy and
 therefore differs from previous calculations using
 the same model at larger $\sNN$\cite{Wei:2018zfb,Li:2017slc,Shi:2017wpk}.
 The difference between the predictions
 made using the 3FD and
 AMPT models becomes larger at
 low collision energy and suggests
 that the polarization is strongly
 dependent on the state of
 the system. We observe rough
 agreement with the calculations made
 using the 3FD model, which
 may imply that the system
 evolves hydrodynamically even at low
 collision energies.
 At a more general level than $\sim1\sigma$
 discrepancies, the observation of large polarization
 demonstrates that the hadron gas supports
 enormous vorticity at low collision energies.

\begin{figure}[t]
  \centering
  \includegraphics[width=\linewidth]{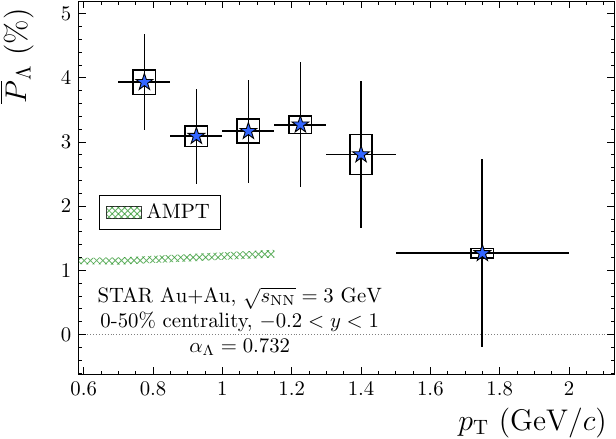}
  \caption{The $\pT$ dependence of $\PLambda$
 is compared with AMPT\cite{guo2021locating} at an
 impact parameter of 8~fm. We
 observe no dependence within uncertainties. Statistical
 uncertainties are represented with lines while
 systematic uncertainties are represented with boxes.}
  \label{fig:Pt}
\end{figure}  

As seen in Fig.~\ref{fig:Centrality}, we observe
 larger hyperon  polarization for more
 peripheral collisions, consistent with the increased
 global angular momentum in the system\cite{Jiang:2016woz}.
 This expectation is borne out by
 the 3FD calculations as well as
 the partonic-transport calculations,
 though the overall scale of the
 latter is much lower than the
 data. A similar dependence of $\PHyper$
 was observed in collisions at two
 orders of magnitude higher energy, $\sNN=200$~GeV\cite{Adam:2018ivw}.
 In Fig.~\ref{fig:Pt}, $\PLambda$ is seen to
 be independent of transverse momentum, within
 uncertainties, similar to the lack of
 dependence seen in top-energy RHIC collisions\cite{Adam:2018ivw}.
 At both $\sNN=3$ and 200~GeV\cite{Adam:2018ivw}, 
 partonic-transport calculations predict only a mild
 dependence.

Global polarization is directly related to
 $\Jsys$, a manifestly three-dimensional phenomenon correlating
 transverse and longitudinal degrees of freedom.
 However, $\PHyper$ decreases with increasing collision
 energy, even as $|\Jsys|$ increases with
 $\sNN$; cf. Fig.~\ref{fig:PolarVsEnergy}. This may be
 partly due to longer evolution times
 at higher energies, increasing the viscosity-driven
 decay of vorticity before polarized hyperon
 emission\cite{Karpenko:2017lyj}. An increased system temperature at
 higher $\sNN$ may also play a
 small role in decreased polarization\cite{Wang:2013xtp}. Several
 models associate the $\sNN$ dependence of
 $\PHyper$ with the vorticity becoming increasingly
 concentrated at forward rapidity, $|y|\gtrsim1\mathrm{-}1.5$, 
 including transport\cite{Jiang:2016woz},
 hydrodynamics\cite{Ivanov:2018eej,Wu:2019eyi,Ivanov:2019ern,Ivanov:2020wak},
 and geometric-driven calculations\cite{Liang:2019pst}. 
 Correspondingly, these
 models predict a strong increase of
 $\PHyper$ as $|y|$ is increased. Still
 other calculations predict a dramatic reduction
 of $\PHyper$ away from mid-rapidity\cite{Deng:2016gyh,Wei:2018zfb,Xie:2019jun}. In
 most models, the dependence becomes stronger
 at lower $\sNN$ since higher-energy collisions
 better approximate boost invariance in the
 mid-rapidity region. 

\begin{figure}[t]
  \centering
  \includegraphics[width=\linewidth]{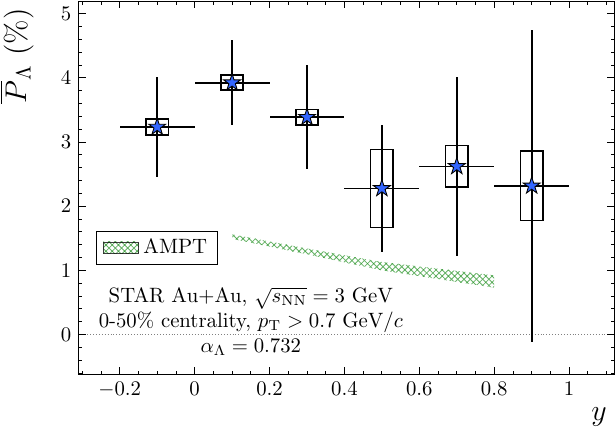}
  \caption{The $y$ dependence of $\PLambda$
 is compared with AMPT\cite{guo2021locating} at an
 impact parameter of 8~fm. We
 observe no dependence within uncertainties, even
 for the most forward $\Lambda$ hyperons.
 Statistical uncertainties are represented with lines
 while systematic uncertainties are represented with
 boxes.}
  \label{fig:y}
\end{figure}

While all previous measurements were confined
 to the region $|y|\ll|y_{\rm beam}|$ and
 were unable to reconstruct forward-rapidity $\Lambda$
 hyperons, the present measurement covers the
 range $-0.2\leq y\lesssim y_{\rm beam}$
 which reaches
 the upper limit of $y_\Lambda$ at
 this collision energy. As shown in
 Fig.~\ref{fig:y}, we find no significant dependence
 of $\PLambda$ on rapidity, though statistical
 uncertainties are relatively large and a
 loose centrality selection is used. This
 is already sufficiently precise to disagree
 with the prediction of AMPT.

Our measurement of nonzero $\PLambda$ at
 $\sNN=3$~GeV demonstrates that vorticity aligned
 with $\JsysHat$ is at a maximum
 below $\sNN=7.7$~GeV. The data agree roughly
 with calculations made using the 3FD
 model, integrated over mid-rapidity, but are
 dramatically larger than such calculations made
 using the partonic-transport model AMPT.
 As in Ref. \cite{Adam:2018ivw}, we observe
 a significant centrality dependence of $\PLambda$
 that is consistent with increasing $\Jsys$.
 Our measurement of the dependence of
 $\PLambda$ on $y$ is uniquely valuable
 because we have access to the
 most forward-rapidity $\Lambda$ hyperons. Interestingly, despite
 the variety of model calculations predicting
 quite strong dependence of $\PHyper$ on
 $y$ \cite{Jiang:2016woz,Ivanov:2018eej,Wu:2019eyi,Ivanov:2019ern,Ivanov:2020wak,Liang:2019pst,Deng:2016gyh,Wei:2018zfb,Xie:2019jun}, 
 we see no statistically
 significant dependence. A migration of $\PHyper$
 towards forward rapidity has been offered
 as a potential explanation of the
 monotonic fall of $\PHyper$ with $\sNN$\cite{Jiang:2016woz}.
 Given our observation, such an explanation
 may be incorrect, though this does
 not dispel such arguments as the
 state of the system at higher
 energy is notably different; measurements of
 $\PHyper$ using the STAR forward upgrade
 will provide indispensable comparisons to the
 work presented here.
  
We thank the RHIC Operations Group
 and RCF at BNL, the NERSC
 Center at LBNL, and the Open
 Science Grid consortium for providing resources
 and support.  This work was
 supported in part by the Office
 of Nuclear Physics within the U.S.
 DOE Office of Science, the U.S.
 National Science Foundation, the Ministry of
 Education and Science of the Russian
 Federation, National Natural Science Foundation of
 China, Chinese Academy of Science, the
 Ministry of Science and Technology of
 China and the Chinese Ministry of
 Education, the Higher Education Sprout Project
 by Ministry of Education at NCKU,
 the National Research Foundation of Korea,
 Czech Science Foundation and Ministry of
 Education, Youth and Sports of the
 Czech Republic, Hungarian National Research, Development
 and Innovation Office, New National Excellency
 Programme of the Hungarian Ministry of
 Human Capacities, Department of Atomic Energy
 and Department of Science and Technology
 of the Government of India, the
 National Science Centre of Poland, the
 Ministry  of Science, Education and
 Sports of the Republic of Croatia,
 RosAtom of Russia and German Bundesministerium
 f\"ur Bildung, Wissenschaft, Forschung and Technologie
 (BMBF), Helmholtz Association, Ministry of Education,
 Culture, Sports, Science, and Technology (MEXT)
 and Japan Society for the Promotion
 of Science (JSPS).

\bibliography{GlobalPolarization3GeV}

\end{document}